\documentclass[pre,showpacs,twocolumn]{revtex4-1}
\usepackage{graphicx}
\usepackage{float}
\usepackage{amsmath}
\usepackage{amsfonts}
\usepackage{caption}
\usepackage{subcaption}
\usepackage[font=small,labelfont=bf]{caption}
\usepackage{hyperref}
\usepackage{url}
\usepackage{color}
\usepackage{array} 

\usepackage{mathtools}
\newcommand{\defeq}{\vcentcolon=}

\newcommand{\TT}{\ensuremath{\mathbb{T}}}
\newcommand{\mean}[1]{\left<#1\right>}
\newcommand{\expectationalpha}[1]{\left<#1\right>_\alpha}
\newcommand{\abs}[1]{\left|#1\right|}
\newcommand{\tw}[1]{{\color{blue}[\kern-2pt [\hbox{\lower.5ex\hbox{\tiny TW}} #1]\kern -2pt]}}

\begin{document}

\author{Jonah A. Eaton}
\email{jonah.eaton@nyu.edu}
\altaffiliation[present address: ]{Physics Department,
New York University, New York, NY 10003, USA}

\author{Brian Moths}
\author{Thomas A. Witten}
\email[corresponding author: ]{t-witten@uchicago.edu}
\affiliation{James Franck Institute, University of
  Chicago, Chicago, IL 60637, USA}
\title{Criterion for noise-induced synchronization: application to colloidal alignment}
\date{\today}

\begin{abstract}
Colloidal bodies of irregular shape rotate as they descend under gravity in solution.  This rotational response provides a means of bringing a dispersion of identical bodies into a synchronized rotation with the same orientation using programmed forcing.  We use the notion of statistical entropy to derive bounds on the rate of synchronization.  These bounds apply generally to dynamical systems with stable periodic motion with a phase $\phi(t)$, when subjected to an impulsive perturbation.  The impulse causes a change of phase expressible as a phase map $\psi(\phi)$.  We derive an upper limit on the average change of entropy $\mean{\Delta H}$ in terms of this phase map; when this limit is negative, alignment must occur.  For systems that have achieved a low entropy, the $\mean{\Delta H}$ approaches this upper limit.   
\end{abstract}

\maketitle

\section{Introduction}
\label{sec:Introduction}

 An important phenomenon in systems of many independent agents is synchronization: Some influence external to each agent induces them to evolve from uncorrelated motion to highly correlated motion. In physics, the synchronization of nuclear spins creates the powerful coherent signals that make magnetic resonance imaging possible \cite{Bloembergen1948}.  In physiology the body's circadian rhythm creates synchrony in numerous somatic processes \cite{pattanayak2014rhythms}.  In the macroscopic world synchronization appears in applause \cite{neda2000physics}, in birth-death cycles of organism populations \cite{ranta1997moran} and in the synchronized firing of neurons \cite{mainen1995reliability}.  One demonstrated means of synchronization is to expose all the agents to a  stochastic disturbance or noise  identical for all the agents \cite{ranta1997moran,teramae2004robustness,nagai2005synchrony}.  Here we show how the rate of synchronization is constrained by simple properties of the noise. We demonstrate these constraints using a recently identified form of synchronization from colloid science.


Usually, synchronization is thought to be achieved through some mutual coupling or common periodic external forces \cite{Pantaleone:2002fk}. However, we focus on the less-considered effect of common non-periodic forces or common noise on synchronization. The synchronization of noninteracting limit cycle oscillators with common noise pulses was first studied by Pikovskii in 1984 \cite{pikovskii1984synchronization}. Using phase-reduction methods, it was later shown that broad classes of randomly driven noninteracting limit-cycle oscillators will synchronize to  a single locally stable limit cycle \cite{jensen1998synchronization} or to a partially synchronized state \cite{nakao2007noise}. 

When one restricts the common external signal to be telegraph noise or impulse noise, the behavior of these non-linear elements can be reduced to a {\em phase map},  relating the phase immediately prior to an impulse to the altered phase long after the impulse, when the system has returned to stable periodic motion \cite{nagai2005synchrony,nakao2005synchrony}. Experiments verified that the functional form of the phase map governs the way small differences in phase between two identical oscillators decay under repeated impulses \cite{nagai2009experimental}.

Here we take a statistical approach to the process of synchronization.  We quantify the rate of progress towards the synchronized state in terms of statistical entropy of a probability distribution \cite{Shannon_1948}.    Given the probability distribution of initial phases of the oscillators, the phase map readily determines the distribution after the impulse.  The statistical approach allows powerful bounds on how the entropy can change as the result of an impulse.  Extending previous work \cite{Moths13}, we show that the average entropy change from an impulse is necessarily more negative than a quantity calculated from the phase map called the ``spreading parameter''.   Moreover, at late stages of synchronization, when the probability is strongly concentrated, we show under weak conditions that the entropy change becomes $equal$ to the spreading parameter, generalizing the results in \cite{nagai2009experimental}.  Previous work recognized that when this parameter was negative, synchronization must occur \cite{nakao2005synchrony}.  The present work quantifies the rate of this synchronization.   

In order to show the utility of these new bounds, we investigate a novel form of noise-induced synchronization arising in colloidal dispersions.  
A colloidal dispersion is an assembly of micron-scale bodies suspended in a liquid.   Within soft matter  physics, there is increasing interest in manipulating colloidal dispersions of identically-made biological or manufactured objects \cite{meng2010,sacanna2011}.  Uses and practical limitations of their rotational response have been much explored recently \cite{Goldfriend:2015wq, Doi05, Andreev10,Makino03}. 

Under gravity these bodies gradually drift downward \cite{Happel65}. Sufficiently irregular bodies respond to constant force by rotating so that a specific axis in the body aligns with the force.  Thereafter these bodies precesses around this axis with a constant angular velocity \cite{Gonzalez04, Krapf09}.   A set of identical bodies in a dilute dispersion rotate together, with random orientations around this axis. This orientation amounts to a phase angle.   By suitable random changes in the direction of forcing, the bodies evolve into a common phase, so that they have a common orientation.  This evolution amounts to noise-induced synchronization.  

In Section \ref{sec:Mathematical} we recall the equation of motion governing the rotation of a colloidal body under external forcing; we then describe a random forcing procedure leading to a phase map.  In Section \ref{sec:entropy} we define the spreading parameter of a phase-map system and derive the bounds on entropy dictated by it, as announced above. In Section \ref{sec:Numerics} we illustrate how altering the forcing of a colloidal system alters the phase map to create various synchronization behaviors consistent with these bounds.  Finally, in Section \ref{sec:Discussion} we discuss limitations of our work and implications for future work.

\section{Rotational response to external force}
\label{sec:Mathematical}

\subsection{Linear response matrix}
\label{sec:TTmatrix}

We consider a rigid body in a fluid with some external force, $\vec{F}(t)$ acting at the center of buoyancy and hydrodynamic drag forces acting on the body's surface \cite{Moths13}. We consider the dynamics in the  ``creeping flow'' regime, in which inertial forces are negligible and the force transmitted to a moving particle by the medium is proportional to the particle's velocity. The hydrodynamic forces $\vec{F}$ and torques $\vec{M}$ are related to the body's velocity $\vec v$ and angular velocity $\vec \omega$ via a proportionality matrix.  This matrix can be represented in dimensionless block form: 

\begin{equation}
\label{eq:model}
\begin{bmatrix} \vec{v} \\ \vec{\omega}R \end{bmatrix} = 
\frac{1}{6\pi \eta R}\begin{pmatrix} \mathbb{A} & \mathbb{T}^T \\ \mathbb{T} & \mathbb{S} \end{pmatrix} 
\begin{bmatrix} \vec{F} \\ \vec{M}/R \end{bmatrix}~.
\end{equation}
Here $\eta$ is the viscosity of the fluid, $R$ is the hydrodynamic radius of the object, and $\mathbb{T}$, $\mathbb{A}$, and $\mathbb{S}$ are  $3\times 3$ sub-matrices \cite{Happel65,Doi05}. These depend on the shape of the body and the position of its center of buoyancy within the body. For simplicity we choose units such that $6\pi \eta$ and $R$ are unity. We choose the center of buoyancy as our origin, thus eliminating the external torque on the body, and we describe the rotational motion of the body by the $3\times 3$ ``twist matrix'' $\mathbb{T}$:

\begin{equation}
\label{eq:twistmatrixdef}
\vec{\omega} = \TT \vec{F}~.
\end{equation}
This $\vec{\omega}$ immediately gives the time derivative of a rotating $\TT$ \cite{Moths13}:

\begin{equation}
\label{eq:diffeq}
\dot{\mathbb{T}} = \left[\left( \TT \vec{F}\right)^{\times},\mathbb{T}\right]~,
\end{equation}
where the brackets denote a commutator and $\vec{v}^{\times}$ is the cross product matrix of a vector $\vec{v}$ with entries given by $\left[\vec{v}^{\times}\right]_{ik} \defeq \epsilon_{ijk} v_j$ \cite{Krapf09}. 

Choosing a constant $\vec{F}$ in the lab frame, we consider the dynamics in a rotating reference frame fixed in the object.  $\TT$ becomes constant and we obtain a differential equation for $\vec{F}$: $\dot{\vec{F}} = -\vec{\omega} \times \vec{F} = \vec{F} \times \mathbb{T}\vec{F}$.

We now consider bodies whose $\TT$ matrices have only one real eigenvector.  We refer to these as ``axially-aligning" bodies as explained below.  Such bodies have one real eigenvalue, denoted $\lambda$. The eigenvector of $\lambda$ defines two opposite directions denoted by unit vectors $\hat\eta$ and $-\hat\eta$.  Evidently, $\dot {\vec F}$ vanishes when $\vec F$ lies along $\pm\hat\eta$: the motion is steady, with constant $\vec\omega$.  The self-aligning property is incompatible with a symmetric $\TT$ (which has three real eigenvalues). Thus a self-aligning $\TT$ must have an antisymmetric part.  The antisymmetric part of $\TT$ depends linearly on the position of the center of buoyancy, governed by the mass distribution within the object \cite{Krapf09}.  There is necessarily a position where this antisymmetric part vanishes; this position is known as the center of twist.  Whenever the center of buoyancy is sufficiently far from the center of twist, two eigenvalues of $\TT$ become complex and the object becomes axially-aligning.  Thus axially-aligning objects form a large class. We shall consider only axially aligning $\TT$ from now on. 

\begin{figure}
\begin{center}
\includegraphics[width=.5\textwidth]{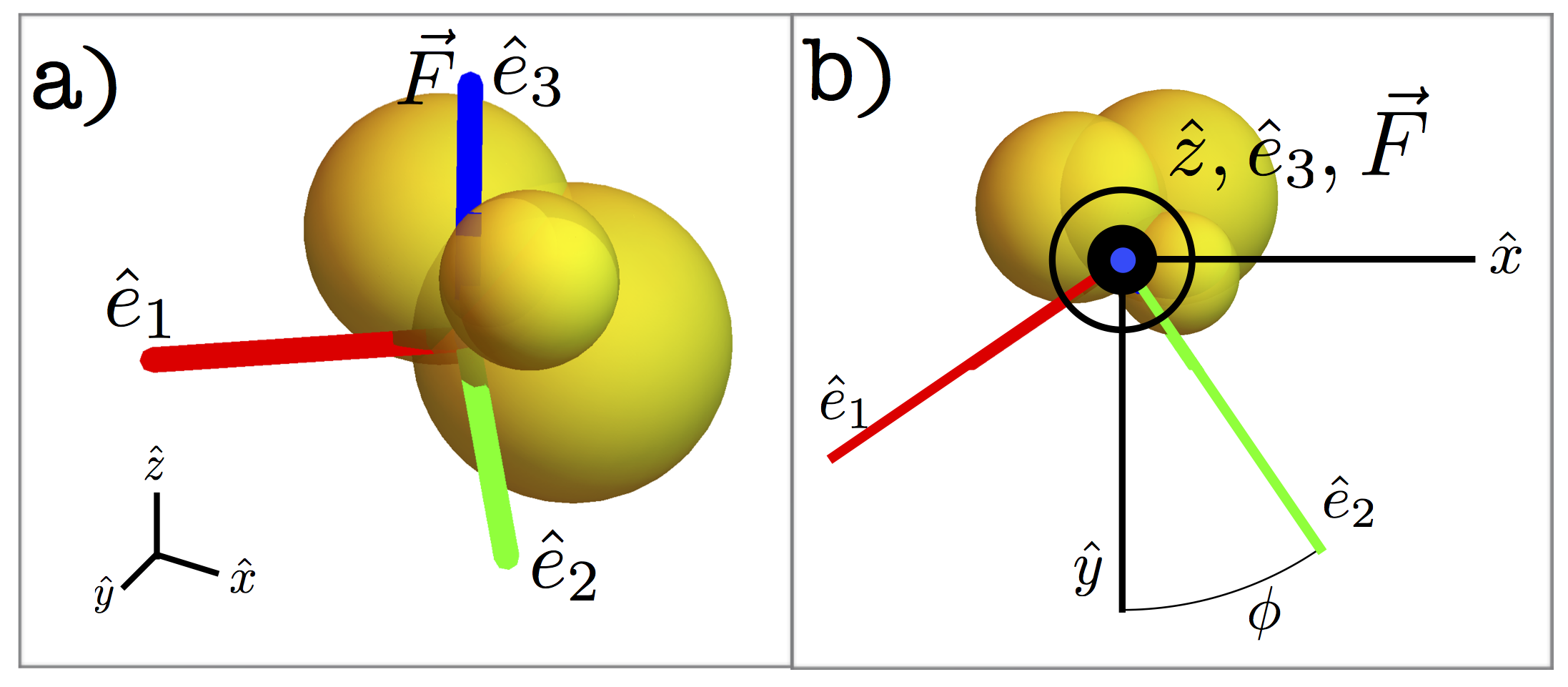}
\captionsetup{justification=raggedright,
singlelinecheck=false
}
\caption{a) An orthographic projection of an axially aligned body with labeled body axes $\hat{e}_1,\hat{e}_2,\hat{e}_3$. b) A top view of the body with the $\phi$ angle being the angle between the $\hat{e}_2$ and the lab $y$-axis. The force is applied in the lab positive $z$-axis as pictured.}
\label{aligns}
\end{center}
\end{figure}

For these $\TT$'s the two orientations $\hat\eta$ and $-\hat\eta$ behave differently.  One of these---denoted $\hat\eta^*$---is a stable steady state.  That is, any initial force direction evolves to the $\hat\eta^*$ direction \footnote{Specifically, the aligning direction $\hat\eta^*$ is one of the two eigendirections $+\hat\eta$ and $-\hat\eta$, ---namely the eigendirection for which $\TT \vec{\delta} \cdot \hat\eta^* < 0$ for any $\vec{\delta} \perp \hat\eta^*$ \cite{Gonzalez04}}.  The  force aligns along the $\hat\eta^*$-axis---hence the name ``axially aligning".  For future use we define a body-fixed basis $\hat{e}_1,\hat{e}_2,\hat{e}_3$ where $\hat{e}_3$ is the aligning direction $\hat\eta^*$ as shown in Figure \ref{aligns}. 

Looked at from the lab frame, a body in a steady state imposed by some constant $\vec{F}$ rotates about this $\vec{F}$ with a constant angular speed given by $\vec{\omega} = \lambda \vec{F}$.  In this frame the stability of the $\hat\eta^*$ direction means that any orientation of the body evolves to make its $\hat e_3$ direction align with $\vec F$, \emph{i.e.} $\vec F \cdot \hat e_3 = |F|$.   (If $\hat e_3$ is initially in the $-\vec F$ direction, the motion is steady but unstable: any slight rotation of the body causes a large rotation of $\hat e_3$ into the stable $+\vec F$ direction.)

Without loss of generality, we assume our force is along the $z$-axis of the laboratory and we now assume the body is in steady state motion. In what follows we define $\tau \defeq 2\pi$ for notational convenience \footnote{further justification of this choice can be found at \url{www.tauday.com/tau-manifesto}}. We then define an azimuthal angle $\phi \in [0,\tau]$ to be the angle between $\hat{e}_2$ and the laboratory's $y$-axis also shown in Figure \ref{aligns}. For any particular body in steady state motion, this $\phi$ increases at the constant rate $\omega$. Given a normalized constant force oriented along the body's $\hat{e}_3$-axis, then $\omega$ is given by:

\begin{equation}
\vec{\omega} = \TT \vec{F} = \lambda \vec{F} = \lambda \\ \hat{e}_3~.
\label{eq:omega}
\end{equation}
The axis of rotation $\hat e_3$ and the azimuthal angle $\phi$ then completely specify the orientation of the body. 

We now consider a dilute dispersion of many such bodies with identical $\TT$ matrices subjected to the same force $\vec F$, but with negligible interactions.  Once a steady state is established, all bodies in the ensemble have a common $\hat e_3$ direction.  They differ only in their $\phi$ angles.  These depend on the history of the sample.

\subsection{Impulsive changes in forcing and phase map}
\label{sec:Forcing}

In this section we first describe a very simple forcing procedure that can be characterized by a phase map.  Our system, when perturbed, returns to its aligned state after some transient period denoted $T$.  In the lab frame, we consider a simple tilt in the applied force by an angle $\theta$ after waiting for a time $t_1  > T$.  This tilted force then acts for a further time $t_2$, also longer than the transient time $T$.  The force $F_\theta(t)$ thus obeys
\begin{equation}
\label{eq:altforce}
\vec{F}_\theta(t) = \begin{cases}
    \hat{z} & \text{for $t \in (0,t_1) $} \\
    \hat{x}\sin\theta + \hat{z}\cos\theta & \text{for $t \in (t_1,t_1+t_2)$}
  \end{cases}~.
\end{equation}
When we switch the direction of the applied force at a time $t_1$, each body rotates to align with the new axis.  After the transient motion, all of the bodies will return to rotating around the tilted $\vec F$. Since the angular velocity is constant and the same for each body in the ensemble, no further alignment of the bodies can be achieved after the transient period. Thus at the final time $t_2$ the bodies again differ from one another by a constant amount in their azimuthal angles, which we denote as $\tilde \phi$.  The complicated transient may cause two similarly oriented bodies to become more similar, or it may make them more different. 

To decide whether the ensemble as a whole is becoming more aligned,  we first make an explicit definition of the azimuthal angles $\phi$ and $\tilde \phi$.  We note that both the new and the old $\vec F$ in  \eqref{eq:altforce} lie in the $x-z$ plane; thus, the $y$-axis is common to both the new and old plane of rotation.  The instant before the switch in forcing angle and the resulting transient motion, we use each body's axis, $\hat{e}_2$ and our lab frame $y$-axis to define $\phi$ for each body in the ensemble. After $\vec{F}_{\theta}$ has switched into the $x-z$ plane and all of the bodies in the ensemble have re-aligned, each body's $\hat{e}_2$ and the lab's $y$-axis again differ by some azimuthal angle $\tilde \phi(t)$.  This allows us to define  a smooth function $\psi_{\theta}(\phi): S^1 \rightarrow S^1$ that maps initial orientations to final orientations.  After the transient $\tilde \phi$ increases linearly in time.  This means that for times $t_2 > T$ $\tilde \phi(t_2)$ can be expressed as a time-dependent term $\omega t$ plus a fixed offset $\psi_\theta(\phi_1)$, where $\phi_1$ is the phase angle at time $t_1$, immediately before the tilt.
\begin{equation}
\tilde\phi(t_2) = \omega~ t_2 + \psi_\theta(\phi_1) ~.
\label{eq:psidef}\end{equation}
This equation defines the phase map $\psi_\theta(\phi)$.  It is evidently the final phase difference extrapolated back to the moment of tilt.  

For $\theta = 0$, there is no change in the applied force and the object maintains its current steady state with no transient motion. With no transient motion, $\psi_0$ is the identity function. Since our differential equations depend smoothly on initial conditions, the deformation of $\psi_\theta$ from $\psi_0$ must also be smooth and $\psi_\theta$ must have a conserved winding number around $S^1$ \cite{Moths13}.  This $\psi_\theta(\phi)$ is a phase map for our system   \cite{nagai2009experimental}.  

This $\psi_\theta$ formulation captures everything important about the dynamics of this system with regard to aligning axially-aligning bodies under the tilted force program defined in \eqref{eq:altforce}.  Using it, we may infer the distribution of phase angles after a single tilt, or after many tilts. 

We can characterize an axially aligned ensemble by a probability distribution $p(\phi)$ which gives the probability of a randomly selected body having the orientation $\phi$ as measured in our lab frame. Then our goal, complete synchronization, corresponds to the probability distribution being a delta function.

\section{Entropy change under a phase map}
\label{sec:entropy} 
The preceding section showed that the effect of an impulsive change in forcing on a colloidal object can be described by the phase map.  Thus in this section we consider an arbitrary dynamical system which, like the colloidal object, has a stable steady state characterized by a phase $\phi$ that increases at a constant rate.  The system may be altered by some sort of impulsive perturbation that changes this phase to $\psi(\phi)$ after the system has returned to a steady state.   Earlier work \cite[p.~95]{Strogatz:2015fk} considered the effect of periodic impulses.  Here we consider the effect of randomly timed impulses \cite{pikovskii1984synchronization}.

We first consider the effect of allowing our ensemble to rotate for a given time. This produces a uniform shift in the initial orientation or phase angle $\phi$ for the entire ensemble and, since there is no transient motion, does not change the overall distribution of phases in the ensemble. The phase $\tilde{\phi}$ after the shift is then given by $\phi + \alpha$ where $\phi$ is the initial phase and $\alpha$ is the size of the shift. The new probability distribution, $\tilde{p}(\tilde{\phi})$,  is merely shifted to the new phase angles, $p(\phi) = \tilde{p}(\tilde{\phi})$. 

Now we consider how the probability distribution transforms under the additional action of $\psi_\theta$, which for simplicity we now denote as $\psi$. The phase after this operation is given by

\begin{equation}
\tilde{\phi} \defeq \psi(\phi + \alpha) ~.
\label{eq:change-variables}
\end{equation}

To diagnose the effectiveness of a $\psi$ to achieve synchronization, we do not attempt to show that the entire circle eventually maps to a single angle. Instead we follow the approach of  \cite{Moths13} and quantify the decrease in randomness of an initially uniform probability distribution $p$. We use information theoretic entropy \cite{Shannon_1948}, $H$, to quantify the disorder of the ensemble. Given some probability distribution function $p$, the functional $H[p]$ is defined as

\begin{equation}
\label{eq:H}
H[p] \defeq - \int p\log\left( p \right) ~,
\end{equation}
where $\log$ is the natural logarithm. We note that as $p$ approaches a delta function distribution, $H[p]$  approaches negative infinity. Additionally, $H[p]$ is maximal when $p$ is constant. 

\subsection{Monotonic $\psi$}
\label{sec:monotonic}
 For a monotonic $\psi$ function, our probability distribution transforms simply:

\begin{equation}
\tilde{p}(\tilde{\phi}) = \frac{p(\phi)}{\psi'(\phi + \alpha)} ~,
\label{eq:monotonic-prob}
\end{equation}
where $\phi = \psi^{-1}(\tilde{\phi}) - \alpha$.  For this case Moths and Witten \cite{Moths13} showed that  on average the entropy must decrease indefinitely with each impulse.

\subsection{Non Monotonic $\psi$}

Moths and Witten showed that it is always possible to choose a $\theta > 0$ small enough that $\psi_\theta$ will be monotonic \cite{Moths13}. However from numerical simulations it was observed that there were non-monotonic $\psi$ that also led to orientational ordering.  Thus we seek a more general condition, valid for non-monotonic $\psi$, that would guarantee an indefinitely decreasing entropy.  Here \eqref{eq:monotonic-prob} no longer applies and a generalized treatment is needed.

\begin{figure}
\begin{center}
\includegraphics[width=.5\textwidth]{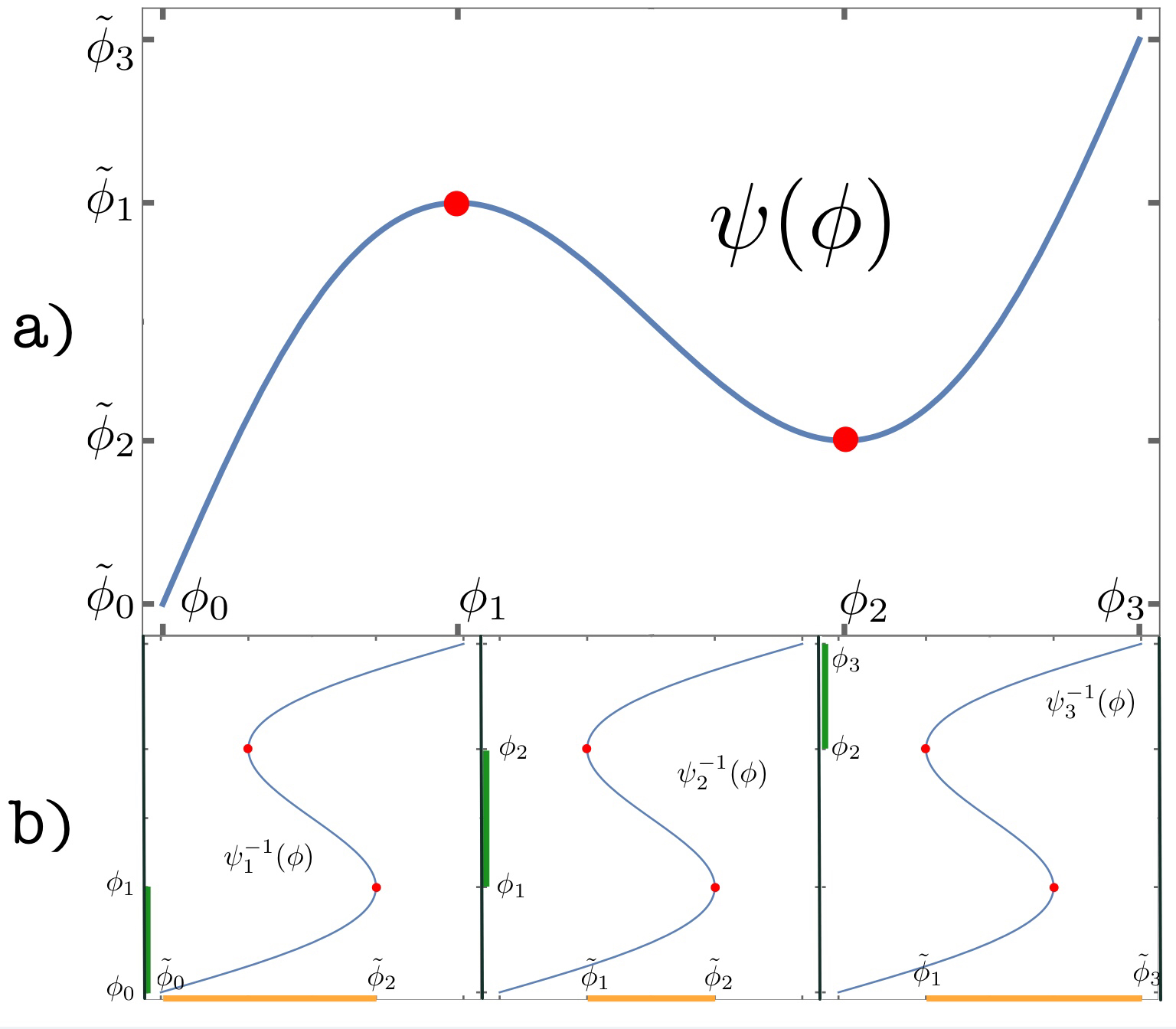}
\captionsetup{justification=raggedright,singlelinecheck=false}
\caption{a) A sketch of a non-monotonic $\psi$ to assist with notation with the extrema shown as (red) dots. b) Sketch of the three inverse functions related to our example $\psi$. We highlight the domain of each inverse in (orange) color along the horizontal axis and the associated target in (green) color along the vertical axis.}
\label{examplePsi}
\end{center}
\end{figure}

We consider some smooth $\psi$ function with a finite number $K-1$ of extrema  such as the one illustrated in Figure \ref{examplePsi} \footnote{There is also the case where $\psi$ is constant over some interval, but we have chosen not to consider as it seems not to occur for $\psi$ functions that arise from \eqref{eq:model} }. (Here we have shifted $\phi$ and $\tilde \phi$ so that $\psi(0) = 0$ and $\psi(\tau)=\tau$).
Since $\psi$ increases by $\tau$ over the range of $\phi$, these extrema divide the domain into $K$ intervals where $\psi(\phi)$ is monotonic. Labeling the extremal $\phi$'s as $\phi_1, ..., \phi_{K-1}$, the $K$ monotonic intervals are then $[0, \phi_1], [\phi_1, \phi_2], ...,  [\phi_{K-1}, \tau] $. Since $\psi$ is strictly monotonic on each interval $[\phi_{k-1},\phi_{k}]$, $\psi$ has $K$ well-defined inverses $\psi^{-1}_k:\tilde{\phi} \rightarrow \phi$. The domain for each $\psi^{-1}_k$ is $[\tilde{\phi}_{k-1},\tilde{\phi}_{k}]$ where $\tilde{\phi}_k = \psi(\phi_{k} + \alpha)$.  An example of such a $\psi$ function with labeled $\phi$ and $\tilde{\phi}$ is shown in Figure \ref{examplePsi}a.

Each monotonic interval in $\phi$ contributes separately to the new probability distribution, $\tilde{p}$, according to the absolute value of \eqref{eq:monotonic-prob}. We denote $\tilde{p}_k$ to be the contribution to the new probability distribution from the $k$th interval. Formally written, 

\begin{equation}
\tilde{p}_k(\tilde{\phi}) = \frac{p(\phi_{(k)})}{\abs{\psi'(\phi_{(k)} + \alpha)}}~,
\label{eq:tilepk}	
\end{equation}
where $\phi_{(k)} = \psi^{-1}_k(\tilde{\phi}) - \alpha$. The $\tilde{p}_k$ vanishes when there is no $k$th pre-image. Summing the contributions from all intervals, we have $\tilde{p} = \sum \tilde{p}_k$. 

With a well-defined probability distribution characterizing how the orientations of the ensemble change, we can now ask whether or not the ensemble becomes more ordered or less ordered.

We now consider the entropy after a transient with shift $\alpha$, $\tilde{H}_{\alpha} = H\left[\tilde{p}\right]$. Our goal is to rewrite $\tilde{H}_{\alpha}$ into the form $\tilde{H}_{\alpha} = H + \Delta H_{\alpha}$ where $H$ is the entropy of the ensemble before the application of $\psi(\phi + \alpha)$. To do so we rewrite $\tilde{H}_{\alpha}$ in terms of the individual contributions $\tilde p_k$ defined above

\begin{equation}
\label{eq:Halpha}
\begin{split}
\tilde{H}_{\alpha} &= - \oint_{S^1} \tilde{p}(\tilde{\phi})\log\left( \tilde{p}(\tilde{\phi}) \right)\mathrm{d}\tilde{\phi} \\
&=  -\oint_{S^1} \left (\sum^{K}_{k=1} \tilde{p}_k(\tilde{\phi})\right ) \log\left( \sum^{K}_{j=1} \tilde{p}_j(\tilde{\phi}) \right)\mathrm{d}\tilde{\phi} ~,
\end{split}
\end{equation}
where each $\tilde{p}_k(\tilde{\phi})$ is given by \eqref{eq:tilepk} for $\tilde{\phi} \in [\tilde{\phi}_{k-1},\tilde{\phi}_{k}]$ and $\tilde{p}_k(\tilde{\phi}) = 0$ elsewhere.

We note that the function $f(t) = t\log(t)$ is a continuous and convex function satisfying the following inequality

\begin{equation}
-\left(\sum^K a_k \right) \log\left(\sum^K a_k \right) \leq -\sum^K a_k \log(a_k) ~,
\label{eq:convexity}
\end{equation}
where each $a_k$ is positive \cite[p.~101]{Rudin}. 

We use this inequality to obtain an upper bound on \eqref{eq:Halpha} which we denote as $H_x$ for simplicity. 

\begin{equation}
\label{eq:inequality}
\tilde{H}_{\alpha} \leq -\oint_{S^1} \sum^{K}_{k=1} \left [\tilde{p}_k(\tilde{\phi}) \log\left( \tilde{p}_k(\tilde{\phi})\right)\right ] ~\mathrm{d}\tilde{\phi} \defeq H_x~.
\end{equation}
 
Since the limits of the sum are independent of $\tilde{\phi}$ we bring it outside of the integral. Additionally each $\tilde{p}_k(\tilde{\phi})$ is non-zero only over a certain interval so the bounds of integration for each integrand can be reduced. 


\begin{equation}
\begin{split}
H_x =& -\sum^{K}_{k=1} \oint_{S^1} \tilde{p}_k(\tilde{\phi}) \log\left( \tilde{p}_k(\tilde{\phi})\right)\mathrm{d}\tilde{\phi} \\
=& -\sum^{K}_{k=1} \mbox{sgn}(\tilde{\phi}_{k-1} - \tilde{\phi}_{k}) \int_{\tilde{\phi}_{k-1}}^{{\tilde{\phi}_{k}}} \tilde{p}_k(\tilde{\phi}) \log\left( \tilde{p}_k(\tilde{\phi})\right)\mathrm{d}\tilde{\phi}~.
\end{split}
\end{equation}

The $\mbox{sgn}$ factor  assures that the limits of integration are in the  conventional increasing order. As we can see in Figure \ref{examplePsi}b, this is not always the case, since for the integral $\int_{\tilde{\phi}_{1}}^{{\tilde{\phi}_2}}$, $\tilde{\phi}_{1}$ is larger than $\tilde{\phi}_{2}$. 

Having separated the integral into separate parts summed together we are in a position to perform a change of variables with $\tilde{\phi} = \psi(\phi + \alpha)$ and then $\mbox{sgn}(\tilde{\phi}_{k-1} - \tilde{\phi}_{k}) \mathrm{d}\tilde{\phi} = \left|\psi'(\phi + \alpha)\right|\mathrm{d}\phi $ and $H_x$ simplifies to

\begin{equation}
\label{eq:H-var-change}
\begin{split}
H_x &= -\sum^{K}_{k=1} \int_{\psi^{-1}_k(\tilde{\phi}_{k-1})-\alpha}^{\psi^{-1}_k({\tilde{\phi}_{k}})-\alpha}\frac{p\left(\phi \right)}{\left|\psi' \left(\phi +\alpha \right) \right|} \\
&\qquad {} \log\left( \frac{p\left(\phi \right)}{\left|\psi' \left(\phi + \alpha \right) \right|}\right) \left|\psi'(\phi + \alpha)\right| \ \mathrm{d}\phi \\
&= -\sum^{K}_{k=1} \int_{\phi_{k-1}}^{\phi_{k}} p\left(\phi \right) \log\left( \frac{p\left(\phi \right)}{\left|\psi' \left(\phi + \alpha \right) \right|}\right) \ \mathrm{d}\phi ~.
\end{split}
\end{equation}

By our construction of the intervals, $[\phi_{k-1},\phi_{k}]$, we can combine our sum over integrals into one integral over the unit circle and then substitute in $H[p]$ using its definition \eqref{eq:H}

\begin{equation*}
\begin{split}
H_x &= -\oint_{S^1} p\left(\phi \right) \log\left( \frac{p\left(\phi \right)}{\left|\psi' \left(\phi + \alpha \right) \right|}\right) \ \mathrm{d}\phi \\
&= H[p] + \oint_{S^1} p\left(\phi \right) \log\left|\psi' (\phi+ \alpha) \right|\ \mathrm{d}\phi ~.
\end{split}
\end{equation*}
Using $\tilde{H}_{\alpha} \leq H_x$ to compare $\tilde{H}_{\alpha} = H[p] + \Delta H_{\alpha}$ with the above we find that $\Delta H_{\alpha} \leq \oint_{S^1}p\left(\phi\right)\log\abs{\psi' (\phi+ \alpha)} d\phi$. Thus to ensure change in entropy $\Delta H_{\alpha} < 0$, it is sufficient to require that $\oint_{S^1}p\left(\phi\right)\log\abs{\psi' (\phi+ \alpha)} d\phi < 0$.

We cannot expect that $\Delta H_{\alpha}$ will be less than zero for all choices of $p$ and $\alpha$. Indeed, if $p$ is concentrated in a region where $\left|\psi'\right| > 1$ our $\Delta H_{\alpha}$ would be positive and the entropy would have increased. 

Though the entropy may increase for particular $p$ and $\alpha$, it need not increase when averaged over $\alpha$. We let the shift $\alpha  \in [0,\tau]$ be chosen randomly and we obtain an upper bound for the expected value of $\Delta H_{\alpha}$, denoted as $\expectationalpha{\Delta H_{\alpha}}$:

\begin{equation}
\label{eq:expvalue}
\begin{split}
\expectationalpha{\Delta H_{\alpha}} & \defeq \frac{1}{\tau} \oint_{S^1} \Delta H_{\alpha} \mathrm{d} \alpha \\
&\leq \frac{1}{\tau}\oint_{S^1}p\left(\phi\right)\mathrm{d}\phi  \ \oint_{S^1} \log\left|\psi' (u) \right|\mathrm{d}u \\
&\leq  \frac{1}{\tau}\oint_{S^1} \log\left|\psi' (u) \right|\mathrm{d}u ~.
\end{split}
\end{equation}

The right side of \eqref{eq:expvalue} is simply the average of $\log\left|\psi' (u) \right|$ over the unit circle, denoted as $\left< \log\left|\psi' (u) \right|\right>$. We define this quantity derived from the phase map as the \emph{spreading parameter}. Previous work \cite{nagai2005synchrony} showed that it is equal to the Lyapunov exponent governing the exponential spreading of two nearby angles under repeated random mappings. Eq. \ref{eq:expvalue} tells us that the bound of the expected change in entropy for a single iteration is independent of the initial probability distribution, $p$ for that iteration. 

Thus far, we have only been considering the change in entropy for a single iteration, but now we wish to consider the change in entropy after many iterations. In general, the average change in entropy after $N$ iterations would be given by $\overline{\Delta H_N} = \frac{1}{N}\sum_{n=1}^N \Delta H_{\alpha_n}\left[p_n\right]$, where $p_n$ is the probability distribution function before the $n$th iteration and $\alpha_n$ is the randomly chosen shift angle at the $n$th step.

To obtain the expected value for the average change in $H$ after $N$ iterations, we take some random sequence of $\left(\alpha_1, \alpha_2, ... \alpha_N\right)$ in the space of $[0,\tau]^N$. We know that while each $p_n$ is dependent on all the previous $\alpha_m$, it is independent of $\alpha_n$. 

\begin{equation}
\label{eq:thm}
\begin{split}
\left< \overline{\Delta H_N} \right> &= \frac{1}{N}\sum_{n=1}^N \left<\Delta H_{\alpha_n}\left[p_n\right]\right>_{\alpha_n}\\
&= \frac{1}{N}\sum_{n=1}^N \left<\Delta H_{\alpha}\right>_\alpha \\
&\leq  \mean{ \log\left|\psi' (u) \right|}_u ~.
\end{split}
\end{equation}

From this we obtain our general constraint for our forcing program to achieve complete alignment: 

\begin{equation*}
\mean{ \log\abs{\psi'}}< 0 ~.
\end{equation*}

If the spreading parameter is negative, then $\left< \overline{\Delta H} \right>$ is guaranteed to be negative as well. When $\left<\overline{\Delta H} \right>$ is negative, the entropy of the system, $H$ will, on average, decrease indefinitely after many iterations. As the number of iterations approaches infinity, $H$ will approach negative infinity. As this occurs, our probability distribution will, on average, be concentrated into a set of zero measure on the unit circle by the central limit theorem \cite{Moths13}.  
When the spreading parameter is negative, it enforces a minimal rate of decrease of the entropy; thus \eqref{eq:thm} gives a gauge of how well a given forcing protocol creates alignment. 

One may now ask whether $\left< \overline{\Delta H} \right>$ can also remain negative when the spreading parameter is positive. Below we argue that it cannot, so that alignment occurs if and only if the spreading parameter is negative.

\subsection{Entropy decrease when entropy is small}
\begin{figure}
\includegraphics[width=\hsize]{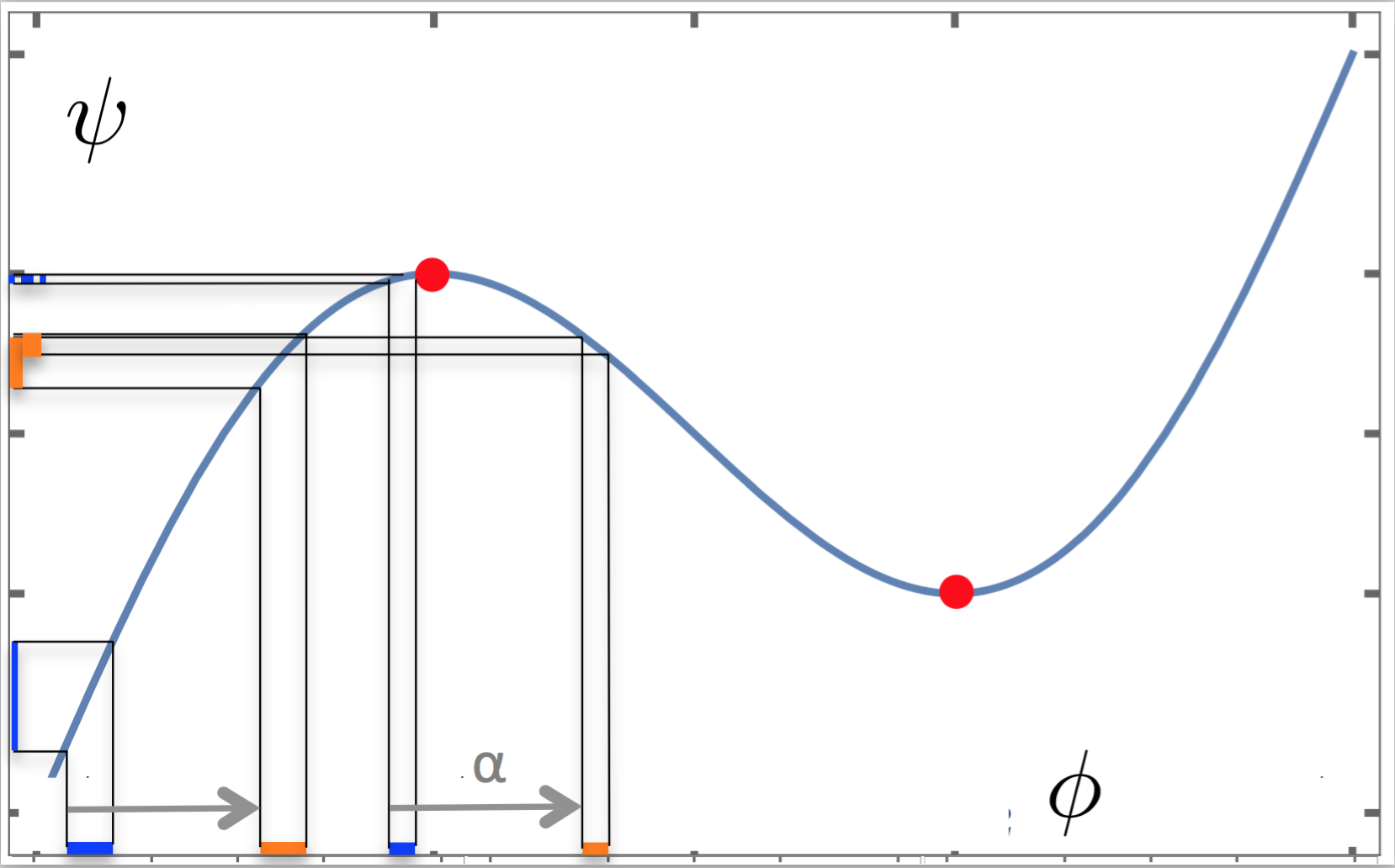}
\captionsetup{justification=raggedright,
singlelinecheck=false
}
\caption{Illustration of multiplicity $C$ when probability is strongly concentrated. Dark colored (blue) bars denote regions of nonzero probability on the $\phi$ axis.  Here each bar maps into a single bar on the vertical $\tilde \phi$ axis and the multiplicity $C(\tilde \phi)$ defined in the text is 1.  The same one-to-one mapping is preserved for generic shift angles $\alpha$.  However, for certain $\alpha$'s, shown by light colored (orange) bars, the mapped regions overlap. The two bars in $\phi$ map into a single bar in $\tilde \phi$.  For this $\alpha$ and $\tilde \phi$ the multiplicity $C(\phi)>1$. As the width of the segments decreases, the fraction of $\tilde \phi$ for which this overlap occurs becomes vanishingly small.}
\label{twosegments}
\end{figure}

When the spreading parameter is negative, the previous section implies that the entropy becomes indefinitely small after many iterations of the force shift.  In this regime we argue that the bound of  \eqref{eq:expvalue}
 limiting $\mean{\Delta H}$ to be less than the spreading parameter becomes an equality.  That is, a new constraint pushes $\mean{\Delta H}$ towards its upper bound.   Indeed, $\mean{\Delta H}$ should approach the spreading parameter as $H[p]\rightarrow - \infty$ even when the spreading parameter is not negative.  The simplification occurs because a small $H[p]$ means that the probability measure $p(\phi)$ is concentrated into an arbitrarily small fraction of the circle.  Our arguments below consider a subset of such $p(\phi)$'s, namely those which vanish except for a finite number of small segments of the circle whose maximum width is $\epsilon$.   Evidently $H \rightarrow -\infty$ as $\epsilon\rightarrow 0$ for such $p(\phi)$'s.

The bound of \eqref{eq:expvalue}
 arises from the convexity property given in  \eqref{eq:convexity}, applied to the final entropy $\tilde H_\alpha$ and the quantity $H_x$.  We now revisit this convexity property for the case where the initial distribution $p(\phi)$ is strongly concentrated.  Now we seek a constraint limiting the separation between $H_\alpha$ and $H_x$.   We may readily choose the $K$ weights $a_k$ in \eqref{eq:convexity} so as to maximize or minimize the difference between the left and right sides of \eqref{eq:convexity}.  We may reduce the difference to zero by choosing all but one of the $a_k$ to vanish.  To maximize the difference, we must fix the sum of the $a_k$, denoted $A$.  Then, the difference is maximal when all the $a_k$ are equal so that $a_k = A/K$ \footnote{To see this, it suffices to note that $\sum a_k \log \sum a_k = A \log A$  and $\sum a_k \log a_k = A \sum (a_k/A)~ (\log (a_k/A) + \log A) = A \log A + A \sum (a_k/A) \log (a_k/A)$. This latter sum is an entropy; as such, it is maximal when all the $a_k$ are equal.}.  
Using this maximum condition we infer 
\begin{equation}
\begin{split}
&\sum^K a_k \log \left (\sum^K a_k \right ) -  \sum^K a_k \log a_k  \\
&\le  A \log A -A \log (A/K)  = A  \log (K) ~.
\end{split}
\label{eq:YY}
\end{equation}

We now apply this relation to $H_x-\tilde H_\alpha$ from \eqref{eq:Halpha} and \eqref{eq:inequality} above.  Using these equations
\begin{equation}
\begin{split}
 H_x -\tilde{H}_{\alpha} = \oint_{S^1} \mathrm{d}\tilde{\phi} & \left [
\sum^{K}_{k=1} \tilde{p}_k(\tilde{\phi}) \log\left( \sum^{K}_{j=1} \tilde{p}_j(\tilde{\phi}) \right) \right. \\
 & \left. -\sum^{K}_{k=1} \tilde{p}_k(\tilde{\phi}) \log\left( \tilde{p}_k(\tilde{\phi})\right) \right ] ~.
\end{split}
\label{eq:ZZ}
\end{equation}

The integrand in \eqref{eq:ZZ} has the same form as the left side of \eqref{eq:YY}, identifying $a_k$ as $\tilde p_k(\tilde \phi)$. We may restrict $\tilde \phi$ to those values with nonvanishing probability $p(\tilde\phi)$, since the integrand vanishes where $\tilde p(\tilde \phi)$ vanishes.   For values of $\tilde \phi$ where $\tilde p(\tilde \phi)$ does not vanish, at least one of the $\tilde p_k(\tilde \phi)$ must contribute.  We call the number of nonzero $\tilde p_k$ the ``multiplicity" for this $\tilde \phi$ and denote it as $C(\tilde \phi)$. Evidently $C$ can be no larger than the number of monotonic regions $K$.  However, for strongly confined initial distributions $p(\phi)$ that vanish over large regions, the multiplicity can easily be smaller than $K$.  Figure \ref{twosegments} shows an example where $K= 3$ but $C=$ 1 or 2. The sums in \eqref{eq:ZZ} have $C(\tilde \phi)$ nonzero terms in them.  Thus we may replace $K$ in \eqref{eq:YY} by $C(\tilde \phi)$.  The $A$ factor is the sum of the $\tilde p_k$ contributions; this is simply $\tilde p(\tilde \phi)$.  Thus the right hand side of \eqref{eq:YY} amounts to $\tilde p(\tilde \phi) \log C(\tilde \phi)$.  Replacing the integrand in \eqref{eq:ZZ} by $\tilde p(\tilde \phi) \log C(\phi)$ yields
\begin{equation}
H_x -\tilde{H}_{\alpha} \le \oint_{S^1} \mathrm{d}\tilde{\phi} ~\tilde p(\tilde \phi) \log C(\tilde\phi) ~. \label{eq:HxminusHalpha}
\end{equation}

As noted in Eq \eqref{eq:H-var-change} $\tilde H_\alpha - H_x$ is simply the difference between the entropy change $\Delta H_\alpha$ and the unaveraged spreading parameter $\oint_{S^1} p\left(\phi \right) \log\left|\psi' (\phi+ \alpha) \right|\ \mathrm{d}\phi$.  Thus the inequality of \eqref{eq:HxminusHalpha} amounts to 
\begin{equation}
\begin{split}
&\oint_{S^1} p\left(\phi \right) \log\left|\psi' (\phi+ \alpha) \right|\ \mathrm{d}\phi -\Delta H_\alpha \\
&\le \oint_{S^1} \mathrm{d}\tilde{\phi} ~\tilde p(\tilde \phi) \log C(\tilde\phi) ~. 
\end{split}
\label{eq:spreadlimit}
\end{equation}
Upon averaging over $\alpha$, the left side becomes $\mean{\log|\psi'|} - \expectationalpha{\Delta H_\alpha}$, which was shown to be greater than zero in \eqref{eq:expvalue}. Combining with \eqref{eq:spreadlimit} we infer
\begin{equation}
0 \le \mean{\log|\psi'|} - \expectationalpha{\Delta H_\alpha} 
\le \expectationalpha{\oint_{S^1} \mathrm{d}\tilde{\phi} ~\tilde p(\tilde \phi) \log C(\tilde\phi) }~. \label{eq:doublesqueeze}
\end{equation}
This inequality is evidently strongest when the multiplicity $C$ is smallest.  We now argue that when $p(\phi)$ is confined to arbitrarily narrow segments, $C$ approaches 1 and the right side of \eqref{eq:doublesqueeze} approaches 0.  Figure \ref{twosegments} shows why narrowing the segments leads to smaller $C(\tilde \phi)$'s.  At a given value of $\tilde \phi$ on the vertical axis, there is typically no probability, and thus no contribution to $C$.  For a small fraction of this axis shown by colored bars, $C$ is defined.  For the typical case, shown in dark color (blue), the probability at every $\tilde \phi$ comes from exactly one bar of nonzero probability in $p(\phi)$.  This $C$ is only greater than 1 in situations like that shown by the light-colored (orange) bars.  Here two different bars of nonzero $p(\phi)$ have mapped into the same $\tilde \phi$ over a small subsegment, via different monotonic intervals of $\psi(\phi)$.  In general $C$ can only be greater than 1 when two or more bars overlap in this way.  

We now reduce the widths of the bars by some common factor.  This has no effect on $\tilde \phi$'s for which  where there was no overlap: $C$ remains 1.  However in cases of overlap like the light bars, the subsegment of overlap evidently decreases.  There is no $\tilde \phi$ for which $C$ increases, and there are overlap regions for which $C$ decreases.  Thus $\tilde p(\tilde \phi) \log C$ must decrease for any normalized distribution $\tilde p$.  There is no bound to this decrease except when $C(\tilde \phi)$ approaches 1 for all $\tilde \phi$.  Thus the right side of \eqref{eq:doublesqueeze} approaches 0 and $\expectationalpha{\Delta H_\alpha}$ must approach $\mean{\log|\psi'|}$.  This reasoning strongly indicates that for generic phase maps $\psi(\phi)$ and generic concentrated $p(\phi)$, the change of entropy must approach $\mean{\log|\psi'|}$ as observed.

\section{Numerical Investigation}
\label{sec:Numerics}

In this section we investigate the effect of our tilting protocol via specific numerical calculations. Our numerical work is of two kinds.  One set of tests is based on integrating  \eqref{eq:twistmatrixdef} through a sequence of tilting forces for a given $\TT$.  A second set of tests infers the final state from the $\psi_{\theta}$ functions of this $\TT$.  We wish to check that (a) the orientational ordering behavior is as expected, (b) whether or not the spreading parameter in \eqref{eq:thm} is a good guide to how the entropy will evolve for a given case.

Our study is conducted via a sequence of four procedures denoted A--D, which we now describe.

\subsection{Creating an ensemble of initial objects}
\label{subsec:Aprog}

\textbf{(A1)} We first generate a $3\times 3$ matrix with entries randomly chosen from the unit interval until a matrix with a complex eigenvalue is found. We designate this to be our original axially-aligning body, represented by $\TT_0$. 

\textbf{(A2)} We apply a constant force along the $z$-axis. The differential equation governing $\TT_0$'s response to a general force in the $x-z$ plane is obtained from \eqref{eq:diffeq}. We record it here for later reference:

\begin{equation}
\dot{\TT}(t) = \left[ \left(\TT \cdot \left(\begin{array}{c}
\sin \theta \\
0 \\
\cos \theta \\
\end{array} \right)
\right)^{\times},\TT \right] ~,
\label{eq:XexplicitDiffEq}
\end{equation}
where $\theta$ is the angle the force makes with the $z$-axis. Thus to find the response of $\TT$ to a force along the $z$-axis, we solve \eqref{eq:XexplicitDiffEq} with $\theta =0$. We solve this differential equation for sufficiently long time $t_{max}$, until the solution's stable real eigenvector, given by $\hat{e}_3$, is properly aligned with the lab's $z$-axis. For simplicity, we designate this properly axially aligned body and the matrix that describes its orientation as $\TT$. 

\textbf{(A3)} We then define the body axis $\hat{e}_2$ such that $\hat{e}_2$ of $\TT$ is parallel the $y$-axis of our lab frame. This gives us a common axis to define our azimuthal angles, $\phi$. Thus for any $\TT$, the angle $-\phi$ is defined as the positive rotation about $\hat{e}_3$ needed to rotate $\hat{e}_2$ into the $y$-axis.

\textbf{(A4.1)} To create an initial ensemble of identical bodies, 500 angles were drawn randomly from $[0,\tau]$, which were used to make 500 different rotations of $\TT$ about the $z$-axis. In the language developed in Section \ref{sec:entropy}, we can think of these 500 angles as our initial azimuthal angles, $\phi_j$, for the bodies in our ensemble $j=1,2,...500$. Notationally, these initial 500 angles, $\{\phi_j\}_0$, designate 500 different angles that correspond to our initial ensemble, $\{ \TT_j \}_0$, where a $\TT_j$ is a $\phi_j$ rotation of $\TT$ about the $z$-axis. We use the subscript zero to indicate that this is the initial ensemble of our iterative scheme. 

\textbf{(A4.2)} Another useful ensemble is one that has nearly a delta function probability distribution. In this case, we proceed as in (A4.1) but we obtain 500 angles drawn randomly from $[0,\frac{\tau}{100000}]$.

\subsection{Determining function $\psi_{\theta}(\phi)$}
\label{subsec:Bprog}

\begin{figure}
\centering
\label{programDiagram}
\includegraphics[width=.95\linewidth]{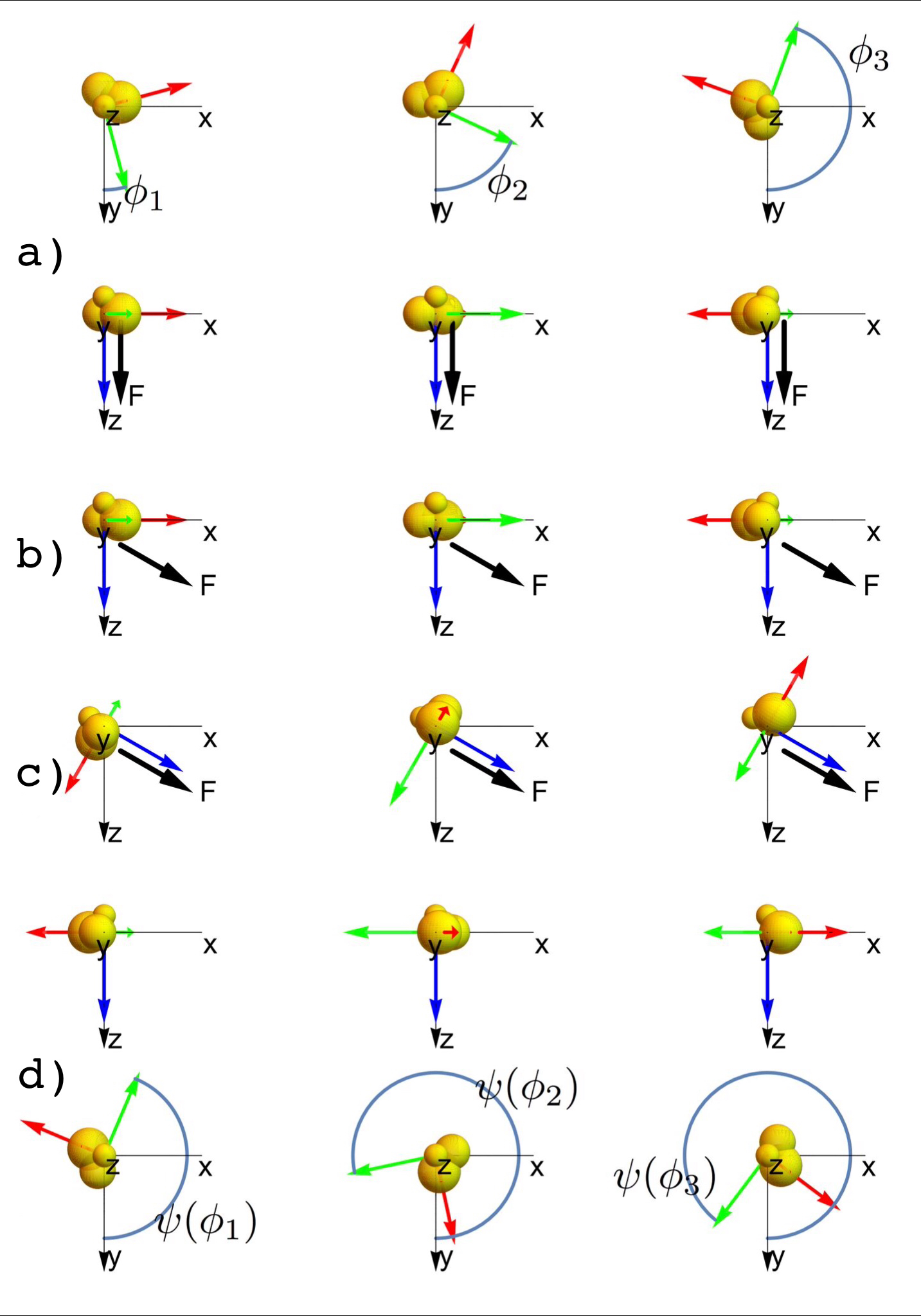}
\captionsetup{justification=raggedright,
singlelinecheck=false
}
\caption{A depiction of the B protocol that is used to compute the $\psi_{\theta}(\phi)$ function.  In the two rows labeled a) we show three representative bodies with from the ensemble $j =$ 1, 2, and 3, as seen from the $z$-axis (top row) and from the $y$-axis (second row). Each body's $\hat{e}_3$ (blue) is already aligned with the lab $z$-axis as enforced by a common force $\vec{F}$. The initial angles $\phi$ are measured from the lab's $y$-axis to the body's $\hat{e}_2$ axis (green). Row b) shows the bodies immediately after $\vec F$ has been tilted. Row c) shows the bodies at some time after they have re-aligned with this $\vec F$.  Rows d) show these bodies rotated non-dynamically so that the aligned direction is again along $z$, in the bottom row, seen from the $z$-axis, the angles $\psi$ are indicated. }
\end{figure}

\textbf{(B1)} We first use \eqref{eq:XexplicitDiffEq} with the nonzero tilt angle $\theta$ to evolve the $\{\TT_j\}$ ensemble from their initial values generated in (A4.1), $\{\TT_j\}_0$ for the time $t_{\max}$.
\textbf{(B2)} The matrix of each body is explicitly rotated about the lab's $y$-axis by $-\theta$, so that each body's $\hat{e}_3$-axis is again parallel to the lab's $z$-axis. \textbf{(B3)} Using the body axis $\hat{e}_2$ defined with $\TT$, we obtain the 500 corresponding values of $\psi$ resulting from the transient motion of (B1). Unlike the construction of $\psi$ function found in \cite{Moths13}, this construction says nothing about what the value of $\psi(0)$ should be\footnote{The construction of $\psi$ found in \cite{Moths13} would guarantee that $\psi(0)=0$. We can recover the previous formulation by either redefining our axis after every application of $\psi$ or by choosing $t_{max}$ such that $\psi(0)=0$ is satisfied.}. \textbf{(B4)} The one-to-one matching of initial $\phi$ to corresponding $\psi$ defines our function $\psi(\phi)$ by interpolation with a 3rd degree polynomial curve between points with periodic boundary conditions.  

\subsection{Evolving the ensemble over many tilts}
\label{subsec:Cprog}

\textbf{(C1)} The forcing program acts on some ensemble, $\{ \TT_j \}_0$ like those constructed in (A4.1) or (A4.2) and the program has $N$ steps, where $N$ is the number of times in our sequence $\{t_n\}$. Each $t_n$ is chosen randomly from $[0,\frac{\tau}{\omega}]$, where $\omega$ is the angular velocity of the body during steady state motion obtained from \eqref{eq:omega}. 

\textbf{(C2)} The $n$th step of the forcing program involves evolving the entire ensemble, $\{ \TT_j \}_{n-1}$, using \eqref{eq:XexplicitDiffEq} with the chosen tilt angle $\theta_n$ for a predetermined, sufficiently long period of time $t_{max}$. We evolve the resulting ensemble further using the same equation for a time $t_n$ taken from $\{t_n\}$. If $n$ is odd, $\theta_n = \theta$, the tilt angle determined in (B1), and when $i$ is even, $\theta_n = 0$. By the rotation at the end of the $n$th step we have obtained the new ensemble $\{ \TT_j \}_{n}$.

\textbf{(C3)} At the end of each step, we take measurements of the ensemble $\{ \TT_j \}_{n}$. For each body that is in the ensemble, $\TT_j$, we find the angle $\phi_j$ about the $\hat{e}_3$ body axis as we did in (B2). This gives us a distribution angles at the $n$th iteration, $\{\phi_j\}_n$. From the distribution we estimate the entropy, $H$ of the system \eqref{eq:H} using a nearest neighbor estimate \cite{victor2002}:

\begin{equation}
H\left[ \{\phi_j\} \right] = \frac{1}{M} \sum_{j=1}^{M} \log\lambda_j + \log\left[2M-2\right] + \gamma ~,
\label{eq:NNestimate}
\end{equation} 
where $\lambda_j$ is the angular distance between $\phi_j$ and its nearest neighbor along the circle, $M=500$ for our work and $\gamma$ is the Euler-Mascheroni constant. 

\begin{figure*}
\centering
\includegraphics[width=.90\linewidth]{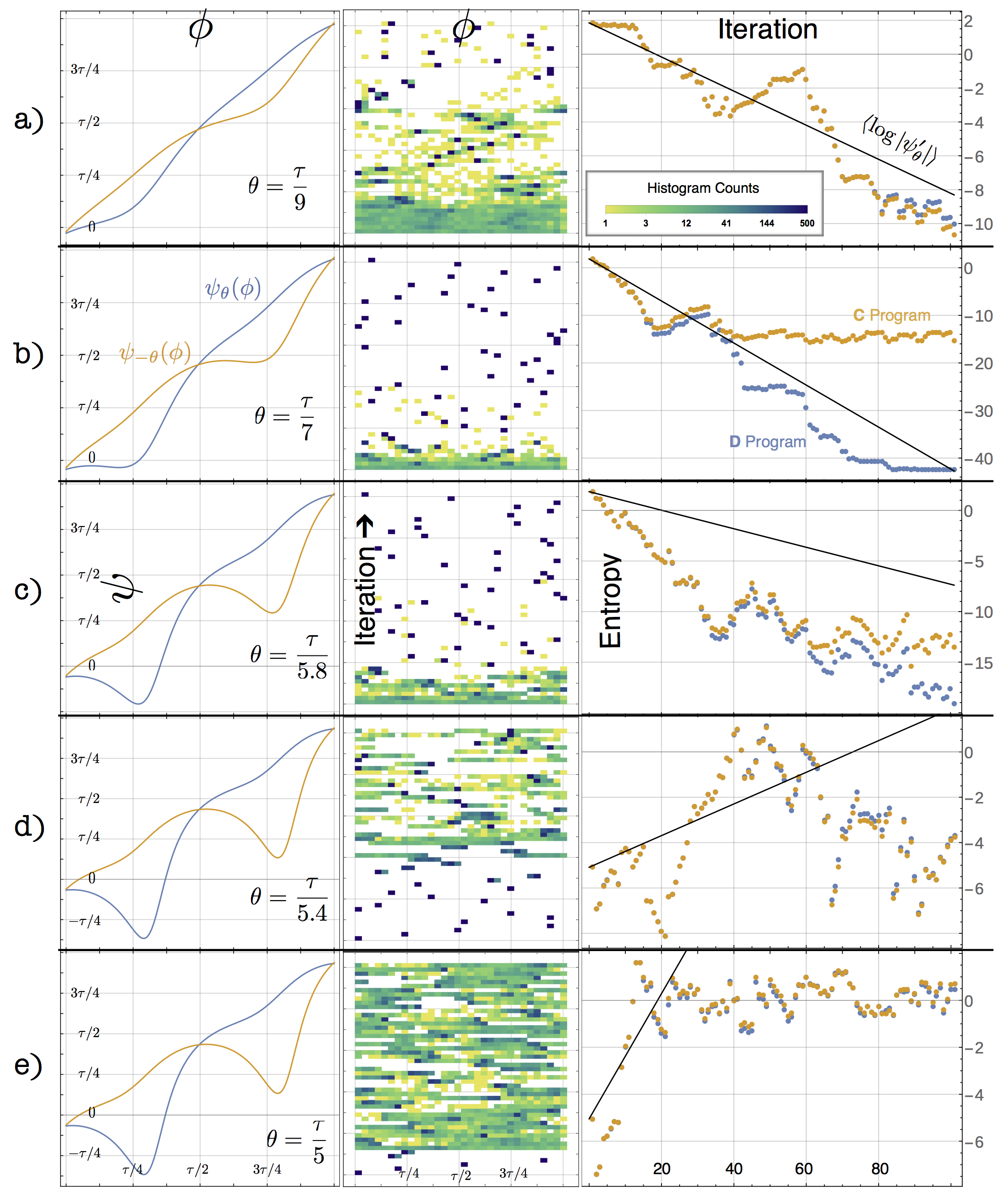}
\captionsetup{justification=raggedright,
singlelinecheck=false
}
\caption{The first column shows the $\psi(\phi)$ functions obtained as described in section \ref{subsec:Bprog}, for both positive and negative $\theta$.  Using the method of sections \ref{subsec:Cprog}, the evolution of $p(\phi)$ is shown in the second column using a density plot with the iteration step of the program increasing upward along the vertical axis.  The inset in Row a) indicates the density scale. As anticipated, the final $p(\phi)$ for a), b), and c) is concentrated near a single angle that jumps discontinuously with each time step.  For rows d) and e), an initially concentrated $p(\phi)$ rapidly spreads to cover a broad range of angles. The third column shows the evolution of the entropy using explicit dynamics of Section \ref{subsec:Cprog} in light color (yellow) and the phase map of Section \ref{subsec:Dprog} in dark color (blue).  The two methods agree except for the smallest entropies.  Here fluctuations due to numerical roundoff error give larger values for the section \ref{subsec:Cprog} method.  
 Straight lines have slope equal to the spreading parameter $\mean{\log|\psi'|}$.  While the sign of the prediction agrees with the behavior in all cases, the rates of decrease agree only qualitatively.  }
\label{numerics}
\end{figure*}

\subsection{Alternative evolution via $\psi$ function}
\label{subsec:Dprog}

\textbf{(D1)} We can also carry out our forcing program without simulating the dynamics at every step. Again, using \eqref{eq:XexplicitDiffEq}, the tilt angle $\theta$ and sequence of times $\{t_n\}$ must first be specified. \textbf{(D2)} We then compute two functions following steps outlined in section \ref{subsec:Bprog}: $\psi_{\theta}(\phi)$ and $\psi_{-\theta}(\phi)$. We note that $\psi_{\theta}(\phi) = \psi_{-\theta}(\phi + \frac{\tau}{2}) - \frac{\tau}{2}$ so all of our analytical arguments remain valid since the derivatives are equal up to a shift in $\phi$. \textbf{(D3)} The program acts on the initial 500 angles, $\{\phi_j\}_0$ that were used to define the ensembles constructed in (A3.1) or (A3.2). As before, the program has $N$ steps, where $N$ is the length of our sequence $\{t_n\}$.

\textbf{(D4)} The $n$th step of the forcing program involves directly applying the function $\psi_{\theta_i}(\phi) + \alpha_i$ modulo $\tau$ to the each azimuthal angle in $\{\phi_j\}_{n-1}$. If $n$ is odd, $\theta_n = \theta$, and when $n$ is even, $\theta_n = -\theta$. Meanwhile $\alpha_n = \omega t_n$, where $t_n$ is the $n$th term in the sequence of randomly chosen times, $\{t_n\}$. At the end of the $n$th step we have obtained the new ensemble of azimuthal angles $\{\phi_j\}_{n}$.

\textbf{(D5)} At the end of each step, we estimate the entropy, \eqref{eq:H} of the angles $\{\phi_j\}_{n}$ that define the ensemble. Again, we use the nearest neighbor estimate \eqref{eq:NNestimate}, this time slightly modified to account for the limits of double-floating point precision so as to avoid indefinite values.

\begin{figure}
\includegraphics[width=.9\linewidth]{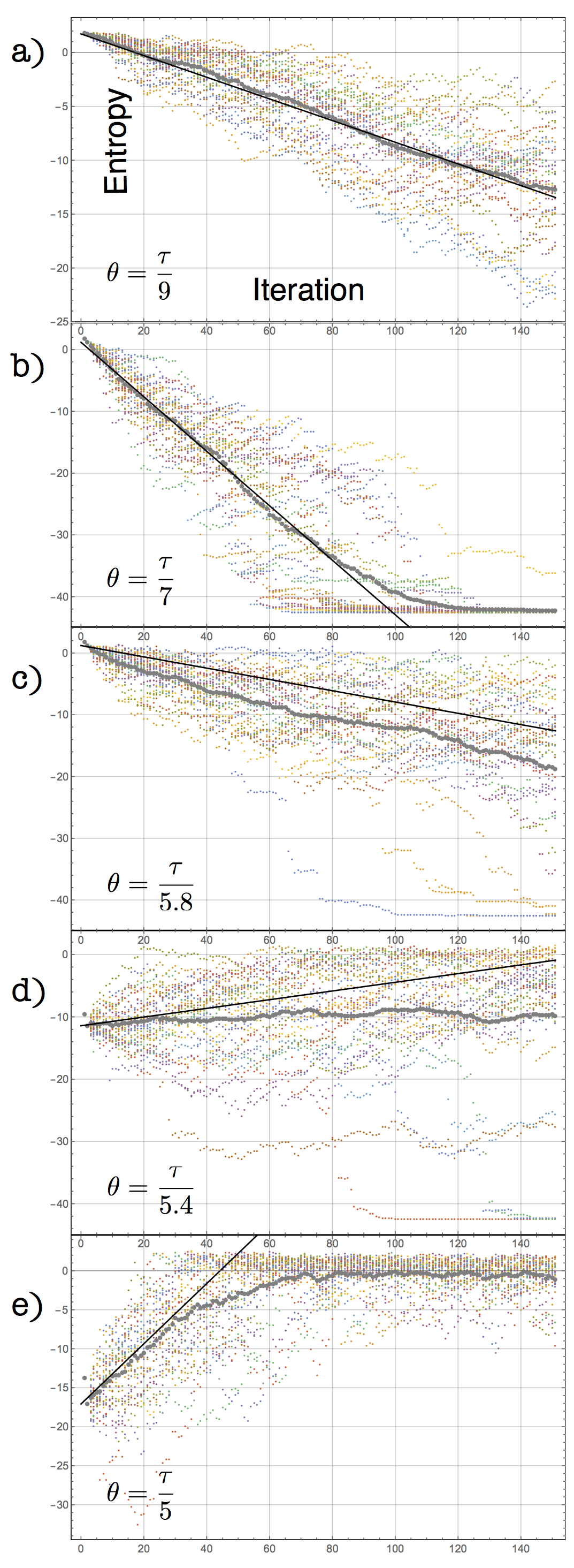}
\captionsetup{justification=raggedright,
singlelinecheck=false
}
\caption{Repeated simulations of the evolution of the entropy using the method found in \ref{subsec:Dprog} using the corresponding $\psi$ functions in Figure \ref{numerics}. Separate runs of the simulations are differentiated by 30 different time sequences, $\{t_n\}$. The average entropy evolution is in dark gray. The solid line has a slope equal to the spreading parameter, the upper bound on the average growth rate.}
\label{multipletrials}
\end{figure}

\subsection{Results}
We performed the simulations described above for several twist matrices $\TT$.  Here  we present the results for a single $\TT$ that was randomly generated \footnote{For all of the sample data depicted in this paper we used 
\begin{equation*} \TT = \left(\begin{array}{ccc}
0.2142 & -0.2350 & 0 \\
0.7634 & -0.5894 & 0 \\
0.2313 & 0.1759 & 0.7133 \\
\end{array} \right) 
\end{equation*}
}.
Figure \ref{numerics} shows the results for five simulations of the forcing program on this $\TT$. For all five simulations we used the same sequence sequence of random times, $\{t_n\}$, so that the angle $\theta$ of the forcing program was the primary differentiator. Additionally, in Figure \ref{multipletrials}, we demonstrate how the alignment process differs for different sequences of random times $\{t_n\}$ as measured by the entropy evolution during each of 30 sequences.  

In left frame of Figure \ref{numerics}a, we have a $\psi$ function that is monotonic, and we would expect the corresponding alternating forcing program will achieve alignment. The center plot is a density plot showing how the initially uniform probability distribution evolves with successive iterations. The right frame shows the evolution of the entropy H with successive iterations.  The solid line has a slope equal to the spreading parameter, $\mean{\log|\psi'|}$, indicating the expected rate of decrease of the entropy.  The entropy indeed fluctuates around this line with a similar average slope.  Figure 6a shows that these fluctuations decrease greatly when one averages the entropy over 30 different simulations. 

Similarly, in the left frames of Figures \ref{numerics}b,c the tilt angles have been increased so that the $\psi$ functions are increasingly non-monotonic. In the center frames of Figure \ref{numerics}b,c, we again show how the probability distribution evolves during 100 iterations. Row b shows a marked increase in the rate of alignment. In the right hand frames the spreading parameter slopes are generally shallower than the observed rates of decrease, illustrating a case when the spreading parameter bound \eqref{eq:thm} is not saturated. Since $\psi$ is non-monotonic, we may only obtain an upper bound the average change in entropy, which is demonstrated for this particular simulation in the right frames of Figure \ref{numerics}b,c. This upper bound relationship on the decrease in entropy is clearer when viewed next to the average of 30 different simulations in Figure \ref{multipletrials}c. While the upper bound appears to be violated in Figure \ref{multipletrials}b, this discrepancy can be attributed to the limits of numerical precision of our simulation when the entropy becomes sufficiently small.

When considering the graphs depicting the change in entropy over many iterations, one will notice that there are periods in which the entropy increases and the alignment is somewhat degraded. One should expect some variability, since there are intervals with $\abs{\psi'} > 1$ that lead to un-alignment as well as intervals that lead to alignment. Since we are choosing a random sequence of times, we would expect that there may be ``unlucky'' parts of that sequence that lead to this variability. 

In the left frames of Figures \ref{numerics}d,e we again have $\psi$ that are not monotonic, but now they have a spreading parameter that is positive, indicating an increase in entropy.  Since we wish to test for such an increase, we generate an initial ensemble that starts out in a nearly synchronized state using (A3.2) for our simulations depicted in the center frames of Figures \ref{numerics}d,e. The spreading parameter is still expected to be an upper bound on the change of entropy and we see in the right frames of Figures \ref{numerics}d,e, that is mostly the case. In view of equation \eqref{eq:doublesqueeze} we also expect that in the beginning of the simulation, while the ensemble is nearly aligned, the spreading parameter should be equal to the change in entropy, which is consistent with the Figures \ref{multipletrials}d,e.

\section{Discussion}
\label{sec:Discussion}

The work presented above broadens understanding of noise-induced synchronization on two fronts.  On the one hand, it provides a quantitative connection between the phase map induced by random impulses and the degree of synchronization it produces.  On the other hand, it illustrates how noise-induced synchronization behaves in the new context of colloidal alignment.  Here we discuss the latter subject first, noting salient features of the numerical experiments, and suggesting implications for colloidal phenomena. We identify the known phenomenon of clustering \cite{nakao2007noise} in relation to our colloidal study and briefly assess the practical applicability of this method.  Next we sketch how our findings involving entropy might prove useful in optimizing noise-induced synchronization in general.  We suggest ways that our entropy-based predictions might be generalized to broader types of noise.

\subsection{Colloidal alignment} \label{sec:colloidal}
As noted above, our numerical results in Figure \ref{multipletrials} on the colloidal system confirm our theoretical claims.  First, the average rates of decrease of the of the entropy were found to be consistent with our spreading parameter, identified as a Lyapunov exponent in prior work.  Second, the decrease became equal to the bound in all cases where the initial entropy was small, as our arguments implied.  Third, the averages predicted by our derivations are well-behaved: one may determine these averages readily using a moderate number of trials.  

In our simulations the predicted average gave useful information about results of a single forcing sequence.  That is, entropies in individual aligning sequences far from the predicted average are rare.  For example, in the system of Figure \ref{multipletrials}b one may predict the number of iterations needed to attain a target entropy of -10.  The average entropy has reached this target in about 17 iterations.  Of the 30 runs contributing to the average, all reached the target in less than 60 iterations, near twice the predicted number. This suggests that the probability of finding entropies greater than -10 after 60 iterations is less than three percent. Similar behavior holds throughout the regime where the observed average follows the predicted average (i.e.\ where numerical errors did not degrade the simulation.)  With high probability the number of iterations needed to attain a given entropy is within a factor 2 of the predicted number.  This statistical regularity seen in our colloidal dynamics suggests that our predicted averages may be similarly useful for synchronization of more general systems.

One aspect of the colloidal system that is ripe for study is the effect of different kinds of  external perturbation or noise.  The noise investigated above was the simplest kind treated in the noise-induced synchronization literature: a sequence of randomly-timed identical impulses.  But since synchronization is observed to occur under much more general noise conditions in the literature, we expect similar generalizations to be possible in our context.  Indeed, our methodology can immediately generalized to the case of impulses of statistically varying amplitude $\theta$.  The effect of changing the amplitude, as seen in Figure 5, is simply to change the $\psi(\phi)$ function.  As shown in the text, any $\psi(\phi)$ that has a negative spreading parameter must reduce the entropy on average.  Thus a random mixture of such impulses must also reduce the entropy indefinitely by the same reasoning that we used for identical impulses.  The literature considers two other aspects of the noise: correlated spacing \cite{hata2010synchronization} and incomplete relaxation between impulses \cite{nakao2007noise, jensen2002synchronization}.  Here too it was found that these generalized noises allowed synchronization.  One is led to speculate that a broad class of random external driving might produce synchronization in our system as well.

This notion leads to an intriguing prospect for a colloidal dispersion.  We imagine that the objects are dispersed in a turbulent fluid, in which each fluid element is undergoing chaotic acceleration.  Locally this acceleration is spatially uniform so that objects within a small region see the same random sequence of accelerations.  As a result one expects nearby objects to become orientationally aligned \footnote{This effect was suggested by Prof. Kevin Mitchel.}.

Achieving synchronized motion in the colloidal system brings practical benefits.  In a colloidal dispersion synchronization means orientational alignment.  With such alignment an anisotropic response such as scattering can provide a new level of information.  The measurement now reflects the properties of the objects at a particular orientation; it shows the effect of different orientations as the objects rotate.  Further, any response that affects the motion of the objects produces the same motion in all of the aligned objects.   This offers ways to manipulate the objects that are not possible without the prior alignment.

Though our system offers a novel case of noise-induced synchronization, our investigation of it has been far from complete.  As noted above, different shapes can lead to a great range of aligning behavior.  This includes bodies that do not have a globally stable aligning direction.  Our study treated only one body as a function of the amplitude ($\theta$) of the perturbations on it.  Nevertheless, prior works \cite{Moths13}  and our own qualitative experience, lead us to believe the synchronization we observed was typical of bodies that self-align along a unique axis.

\subsection{Broader implications}\label{sec:broader}

Our experience with the colloidal system illustrates both the benefits and the limitations of our entropic approach.

Our simulations showed interesting behavior even when the spreading parameter was positive.  Here when the initial state had small entropy, the average rate of increase agreed with the spreading parameter.  However this increase crosses over to a state of constant entropy indicating partial order.  This constant appears to increase as the spreading parameter increases.  Similar behavior has been noted in the noise-induced synchronization literature \cite{nakao2007noise}. There the concentration of the phase angles into a few narrow intervals is known as ``clustering."  It appears that the entropy language may be a useful way to quantify this clustering.

An intrinsic limitation of our approach lies in the use of the entropy measure.  The entropy measures concentration of probability to a small set of phase angles, but a low entropy does not entail synchronization to a single phase angle.  Despite this limitation, the entropy does give a valuable measure of synchronization. Further, in our colloidal simulations, in all the cases where the entropy decreased to a numerically limited level, the final state had converged to a single narrow region of phases. 

The noise-induced synchronization literature will be a valuable guide in generalizing from the simple noise treated here to general noise \cite{pikovskii1984synchronization,jensen1998synchronization,teramae2004robustness,nakao2007noise}.  Likewise, our methods may prove useful for the broader class of noise induced synchronization phenomena.  For example, our methods can be useful in optimizing the amplitude of the noise.  For any given oscillator and noise source, including our colloidal examples, there is a well-defined optimal amplitude.  When the noise is too weak, synchronization is slow; when the noise is too strong the noise opposes synchronization rather than promoting it.  Our methods indicate that the optimum amplitude as measured by entropy is that which gives the most negative spreading parameter. In the literature our spreading parameter is identified as a Lyapunov exponent.  There the notion that more negative Lyapunov exponents means better synchronization is well recognized.  Our main contribution is to show that one may make this relationship quantitative by using the entropy measure.  

\section{Conclusion}

A significant class of colloidal dispersions can in principle be aligned by impulse noise.  The potential benefits of this alignment are great, as noted above.  Yet the experimental feasibility of gaining these benefits has yet to be explored. This work provides a new path to understand, optimize and generalize this type of alignment.  To develop these methods seems promising for further study.  

Additionally we have shown how entropy may be used to study the rate of synchronization in more general systems.  Since the behavior of entropy can be related to simple quantities related to the intrinsic, deterministic dynamics, it appears to be a useful way to characterize future realizations of noise-induced synchronization. 

\section*{Acknowledgements} We are grateful to Amy Kolan for leading us to the noise-induced synchronization literature and for many manuscript improvements.  Arvind Murugan and Matt Ingalls offered further improvements.   We thank our collaborators Haim Diamant and Tomer Goldfriend for fruitful discussions.  Shankar Venkataramani and Kevin Mitchel provided further insights.  This work was supported in part by a grant from the US-Israel Binational Science Foundation and by the National Science Foundation's MRSEC Program under Award Number DMR-1420709.
J.E. thanks the Physics Department of New York University and the Zidovska Lab for support, and T. W. thanks the Kavli Institute for Theoretical Physics for hospitality during the completion of this work.

\bibliography{savedrecs}{}



\end{document}